\documentclass[a4paper,10pt]{article}
\usepackage{graphics}
\usepackage{graphicx}
\usepackage{amssymb}
\usepackage{amsmath}

\title{The spin glass-antiferromagnetism 
competition in Kondo-lattice systems in the presence of a 
transverse applied magnetic field}
\author{S.G. Magalhaes\footnote{ggarcia@ccne.ufsm.br},
F.M. Zimmer
\\
{\it Departamento de F\'{\i}sica, Universidade Federal de Santa Maria,}
\\ 
{\it 97105-900 Santa Maria, RS, Brazil}
\\
\\
B. Coqblin
\\
{\it Laboratoire de Physique des Solides, Universit\'e Paris-Sud, }
\\{\it 91405 Orsay, France}}

\begin{document}
\textwidth=18cm
\textheight=23cm
\date{}
\maketitle

\begin{abstract}
A theory is proposed to describe the competition among 
antiferromagnetism (AF), spin glass (SG)
and Kondo effect. The model describes two Kondo sublattices with an 
intrasite Kondo interaction strength $J_{K}$
and a random Gaussian interlattice 
interaction in the presence of a transverse
field $\Gamma$. 
The $\Gamma$ field is
introduced as a quantum mechanism to produce spin flipping and the random coupling
has average $-2J_0/N$ and variance $32 J^{2}/N$. The path 
integral formalism with Grassmann fields is used to study this fermionic problem,   
in which the disorder is treated within the framework of the replica trick. 
The free energy and the order parameters are obtained using the static ansatz. 
In this many parameters problem,
we choose $J_0/J \approx (J_{K}/J)^{2}$ and $\Gamma/J \approx (J_{K}/J)^{2}$ 
to allow a better comparison with the experimental findings.
The obtained phase diagram has not only the same
sequence as the experimental one for $Ce_{2}Au_{1-x}Co_{x}Si_{3}$, 
but mainly, it also shows a qualitative agreement concerning the behavior 
of the freezing temperature and the Neel temperature 
which decreases until a Quantum Critical Point (QCP).
\end{abstract}

The competition between RKKY interaction and Kondo effect has a fundamental role in $Ce$ and $U$ compounds \cite{BCoqblin}. The presence of disorder in alloys can deeply affect 
such competition and, therefore, it can lead a quite intriguing issue.   
For instance, the $CeAu_{1-x}Co_{x}Si_{3}$ alloy has a phase diagram which displays the 
sequence of phases spin glass (SG), antiferromagnetism (AF) and a Kondo state when 
 the {\it chemical disorder is increased} by substituting $Co$ in the cited alloy \cite{Majumdar}. Moreover,
the Neel temperature 
decreases until reaching a Quantum Critical Point (QCP) at some 
particular value of the $Co$ content, with no evidence of Non-Fermi Liquid behaviour. 
 
Earlier theoretical effort \cite{Magal1} has studied the competition among spin glass, antiferromagnetism and Kondo effect based on a framework previously introduced to study 
the presence of SG in disordered Kondo lattice \cite{Alba}. 
The problem has been treated in a mean field level using functional integral formalism. 
As most important result, 
 a phase diagram which
reproduces exactly the experimental sequence of phases of the $CeAu_{1-x}Co_{x}Si_{3}$ has been obtained. 
Nevertheless, this approach has a fundamental shortcoming, the Neel temperature displays a 
behaviour enterely distinct from the experimental one, it {\it does not} decrease towards a QCP.  
The problem is that the model proposed in reference \cite{Magal1} lacks a quantum mechanism  
able to produce spin flipping. 

In the present work, this mechanism has been incorporated by adding a transverse 
field $\Gamma$ to the original model given in Ref. \cite{Magal1}. 
Recently, this approach has been successfully used to study the SG solution 
in Kondo lattice \cite{AlbaCoqblin}
or in competition with AF \cite{Zimmer2}.   
Therefore, the Hamiltoniam is given by
\begin{align}
H-\mu N= \sum_{p=A,B}[\sum_{i,j}\sum_{\sigma=\uparrow \downarrow}
t_{i j}\hat{d}_{i,p,\sigma}^{\dagger}\hat{d}_{j,p,\sigma} +
\nonumber\\
\sum_{i}\varepsilon_{0, p}^{f}\hat{n}_{i,p}^{f}
+J_{K}\sum_{i}(\hat{S}_{i,p}^{+}\hat{s}_{i,p}^{-}+
\hat{S}_{i,p}^{-}\hat{s}_{i,p}^{+})]
\nonumber\\
+\sum_{i, j} J_{i j}\hat{S}_{i,A}^{z}\hat{S}_{j,B}^{z}+
2\Gamma\sum_{i} (\hat{S}_{i,A}^{x}+\hat{S}_{i,B}^{x})
\label{e1}
\end{align}
where $i$ and $j$ sums over $N$ sites of each sublattice. The intersite coupling 
$J_{ij}$ is a random variable following a Gaussian distribution with average $-2J_0/N$ 
and variance $32 J^{2}/N$. The spin operators are defined as in references \cite{Magal1,AlbaCoqblin}.
The model in Eq.(\ref{e1})
has two-Kondo sublattices with a random Gaussian coupling among localized spins 
only in distinct sublattices \cite{Magal1}. On the other hand, the hopping of the conduction electrons 
is allowed only inside the same sublattice \cite{Magal1}.  

The partition function is treated in the fermionic path integral formalism where the spin operators are
represented by Grassmann fields. The free energy is obtained in the static approximation 
and the replica method is used to average over the random couplings $J_{ij}$.
The fundamental tool in the present case, which has allowed us to  calculate the partition function, consists 
in introducing a matrix formalism with a proper mixing of spinors of each sublattice. Further details will be shown elsewhere \cite{SolCom}.
\begin{figure}[!ht]
\begin{center}
\includegraphics[angle=270,width=10.cm]{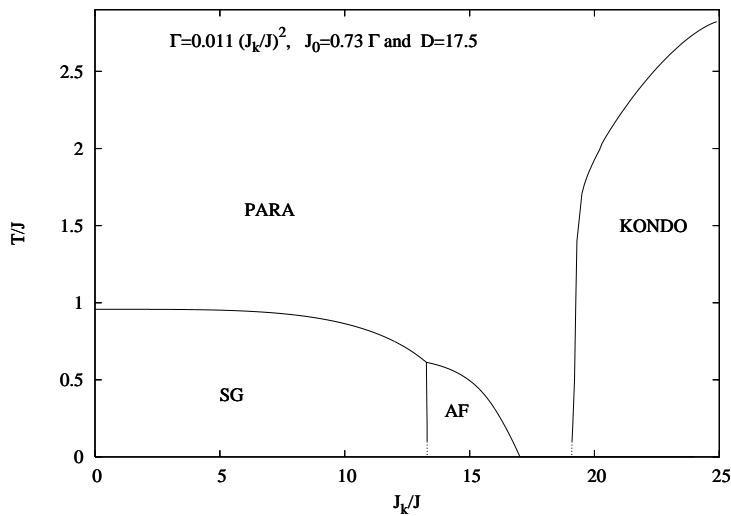}
\end{center}
\caption{Phase diagram $T/J$ {\it versus} $J_{K}/J$ for the relations $\Gamma=
0.011 (J_{K}/J)^{2}$ and $J_{0}/J=0.73 \Gamma/J$. The dotted lines are the
extrapolations carried for lower temperature.}
\label{fig1}
\end{figure}

The free energy is obtained as:
\begin{align}
\beta F= \sum_{p}[\beta J_K|\lambda_p|^2 +\frac{\beta^2 J^2\bar{\chi}_p}{2}(
\frac{\bar{\chi}_{p^{'}}}{2}+q_{p^{'}})
\nonumber\\
-\frac{\beta J_0}{2} m_p m_{p^{'}}-
\frac{1}{2} \int_{-\infty}^{\infty}D\xi_{p}
\ln\int_{-\infty}^{\infty}D z_{p}e^{E( H_{p})}]
\label{e28}
\end{align}
where
\begin{equation}
E(H_{p})=\int^{+\beta D}_{-\beta D}\frac{dx}{\beta D}\ln\{\cosh\frac{x+H_{p}}{2}+
\cosh\sqrt{\Delta} \}
\label{Ep}
\end{equation}
with
$\Delta=[(x-\beta H_{p})^{2}/4+(\beta J_{K} \lambda_{p})^{2}]$,
$H_{p}=\beta\sqrt{\Gamma^{2}+h_{p}^{2}}$,  the internal field 
$h_{p}=J\sqrt{2q_{p'}}\xi_{i,p}+J\sqrt{2\overline{\chi}_{p'}}z^{\alpha}_{i,p}- J_{0}m_{p'}$
($p\neq p^{'}$) and $Dz=e^{-z^{2}/2}/\sqrt{2\pi}$. The saddle point equations for $m_{p}$ (sublattice magnetization), 
$q_{p}$ (SG order parameter) and $|\lambda_{p}|$ (Kondo order parameter) follow directly from
equations (\ref{e28})-(\ref{Ep}). In the present fermionic formulation, the static 
susceptibility  
$\chi_{p}=\beta\bar{\chi}_p$ is an additional saddle point order parameter to be solved with 
previous ones.

One important point in the present approach is to assume the conjecture that the 
parameters $J_{0}/J \approx (J_{K}/J)^{2}$ \cite{Magal1} and $\Gamma/J \approx (J_{K}/J)^{2}$ \cite{AlbaCoqblin}. These 
relationships are introduced to mimic the relation between the Kondo and RKKY interactions. 
Therefore, an one free parameter ($J_k/J$) theory can be build 
where the order parameters solutions are shown in the phase diagram $T/J$ 
($T$ is the temperature) {\it versus} $J_{K}/J$ (see Fig. (1)) with 
AF ($m_{A}=-m_{B}\neq 0$), SG ($q_{A}=q_{B}\neq 0$) and Kondo state ($\lambda_{A}=\lambda_{B}\neq 0$) 
solutions. As one can see in Fig. (1), as long as $J_{K}/J$ increases, it  
is found first a SG, and then an AF solution. The emergence 
of these particular solutions is basically controled by the relationship $J_{0}/J$ \cite{Zimmer2}. In that range of $J_{K}/J$, $\Gamma/J$ is also increasing from zero, which implies that the freezing temperature (the SG transition temperature) has a slight decreasing. However, when AF solution appears, $\Gamma/J$ is strong enough to supress magnetic order leading the Neel temperature to a QCP. From now on, the increasing 
of $J_{K}/J$ can only produce a Kondo state.  
If the $Co$ content can be associated with the 
$J_{K}/J$,  the results shown in Fig. 1  reproduce the basic aspects of the experimental $CeAu_{1-x}Co_{x}Si_{3}$ phase diagram concerning the onset of phases and the behavior of the temperature transitions. 

To conclude, we present here a mean field theory which has a proper set of 
parameters ($J_{0}/J$, $J_{K}/J$, $\Gamma/J$) in the sense that they 
can be related in order 
to capture the essential effects of the disorder in the competition between 
Kondo and RKKY interactions for the $CeAu_{1-x}Co_{x}Si_{3}$ alloy.    

\section*{Acknowledgement}
This work was partially supported by the Brazilian agency CNPq (Con\-se\-lho Na\-cio\-nal de De\-sen\-vol\-vi\-men\-to 
Ci\-en\-t\'\i\-fi\-co e Tec\-no\-l\'o\-gi\-co).

\end{document}